# ANALYTICAL AND EXPERIMENTAL STUDY OF CONICAL TELESCOPING SPRINGS WITH NON-CONSTANT PITCH

**Manuel Paredes**
Université de Toulouse; INSA, UPS, Mines Albi, ISAE; ICA (Institut Clément Ader)
135, avenue de Rangueil,
F-31077 Toulouse, France
manuel.paredes@insa-toulouse.fr

## ABSTRACT

Most research papers that exploit conical springs focus only on conical springs with a constant pitch. In order to increase the range of possibilities for designers, this paper proposes a study of conical springs with other types of spirals projected on the conical shape. This study is related to three other types of conical springs: with a constant helix angle, with a constant stress at solid and with a fully linear load-length relation. For each spring, we give the equation of the spiral, the formula of the initial stiffness, and formulae to calculate the non-linear part of the load-length relation for fully telescoping springs. We also report an experimental study performed to analyze the accuracy of the proposed study based on springs made by fused deposition modeling.

## NOMENCLATURE

$\alpha$ current helix angle
$\theta$ angle that defines the position on the conical helix
$\delta$ deflection
$\delta_1$ deflection of the part of the spring that is free to deflect (from $D_1$ to $D_L$)
$\delta_2$ deflection of the part that is at solid (from $D_L$ to $D_2$)
$\tau$ uncorrected stress
$\tau_i$ constant uncorrected stress at solid.
d wire diameter
dn elementary coil
dr elementary radial displacement on the conical shape
dl elementary orthoradial displacement on the conical shape
dz elementary axial displacement on the conical shape
$D_1$ minimum mean diameter
$D_2$ maximum mean diameter
$D_L$ diameter that defines the limit between coils that are free to deflect and coils at solid for a given load P
F initial spring flexibility
$F_e$ elementary flexibility
G torsional modulus
La active length
$L_f$ free length
$N_a$ number of active coils
$N_L$ number of active coils that defines the limit between coils that are free to deflect and coils at solid for a given load P
$p_e$ elementary axial pitch
P axial load
$P_M$ load when all the coils come to solid (when $D_L=D_1$).
$P_T$ load from which the spring starts to come to solid (when $D_L=D_2$)
r running helix radius (varies from $D_1/2$ to R)
R current helix radius (varies between $D_1/2$ and $D_2/2$)

## 1. INTRODUCTION

Springs are often used in mechanical devices for their ability to store and return energy. The range of applications is very wide. As recent examples, springs can be found in gravity equilibrators [1], laser light applications [2] or non-linear joints of robots [3]. Designers have various data available to help them. A common reference book related to spring design was written several years ago by Wahl [4]. Since that time, many works have improved our knowledge of springs. Recent works focus on testing new materials, such as springs made of memory alloys [6-7], composite springs [8] or nanoscale springs [9]. Other works are related to finding new spring geometries, such as for Belleville springs [10-11], wave springs [12] or non-linear springs [13-14]. Some works also try to improve knowledge of common springs. Dym [15] proposes extended formulae to calculate the rate of cylindrical helical springs. Finally, the dynamic aspect of springs has often been studied for cylindrical [16-17] and non-cylindrical springs [18-20].

In some applications, conical springs can be preferred to cylindrical springs for their ability to have a smaller solid length. Depending on their geometry, they may be able to fully telescope, inducing a solid length equal to the wire diameter. Moreover, they can offer a non-linear load/length curve and this property can be exploited in some applications, such as vibration absorbers [21]. Wu et al [22] have proposed an approximate load-length relation for conical springs. More recently, Rodriguez [23] proposed an analytical law to fully describe the linear and non-linear load-length curve of conical springs with constant pitch. Analytical formulae can be useful for designers, especially in preliminary design when many possibilities have to be evaluated. Analytical formulae are also useful to build high-level assistance tools where metamodels are exploited. For example, the formulae describing the static behavior of conical springs with constant pitch have been used to define an assistance tool dedicated to the optimal design of conical springs [24] and are also included in industrial software proposed by the Institute of Spring Technology [25] or by the Spring Manufacturers Institute [26].

Thus, conical springs with constant pitch are quite well known but many other types of conical springs can be manufactured. For a given conical shape and a given number of turns, the properties of the spring can evolve significantly depending on the way the coils are distributed along the conical profile. In fact, cylindrical springs can simultaneously possess a constant pitch, a constant angle, a constant stress at solid and a linear behavior whereas, for a conical spring, only one property can be achieved at a time. As far as we are aware, there is a lack of analytical description of the geometry or static behavior of conical springs with non-constant pitch and the work reported here is intended to fill that gap.

The previous analytical studies on conical springs used the fact that the solid length geometry was known. In order to be able to develop the same approach, this paper focuses on the study of conical springs that fully telescope so that, however the coils are distributed along the conical shape, the final geometry is known as the spring becomes completely flat. The paper first presents the parameters used to describe the spring helix and the reference formulae to define the spring behavior. These formulae are then used to study a conical spring with a constant angle in section 3. The next section is dedicated to the study of a conical spring that leads to a constant stress when fully telescoped. Section 5 treats a conical spring with a fully linear load-length behavior. A case study is presented in section 6. This section includes an experimental study to test the accuracy of the analytical formulae proposed. The conclusions are drawn in section 7.

## 2. DESCRIPTION OF THEORETICAL MODELS

### 2.1 PARAMETERS OF A CONICAL SPRING

A conical compression spring made of circular wire is studied. Its design can be defined by six parameters (see Fig. 1).

$N_a$ represents the active coils of the spring. In order to make the external load as close to an axial load as possible, two end coils are added, one at the top and one at the bottom of the spring. When end coils are correctly defined, they do not influence the behavior of the active coils. For this reason, the present work refers essentially to the active coils of the spring. At this stage, the way the active coils are distributed along the conical shape is not described.

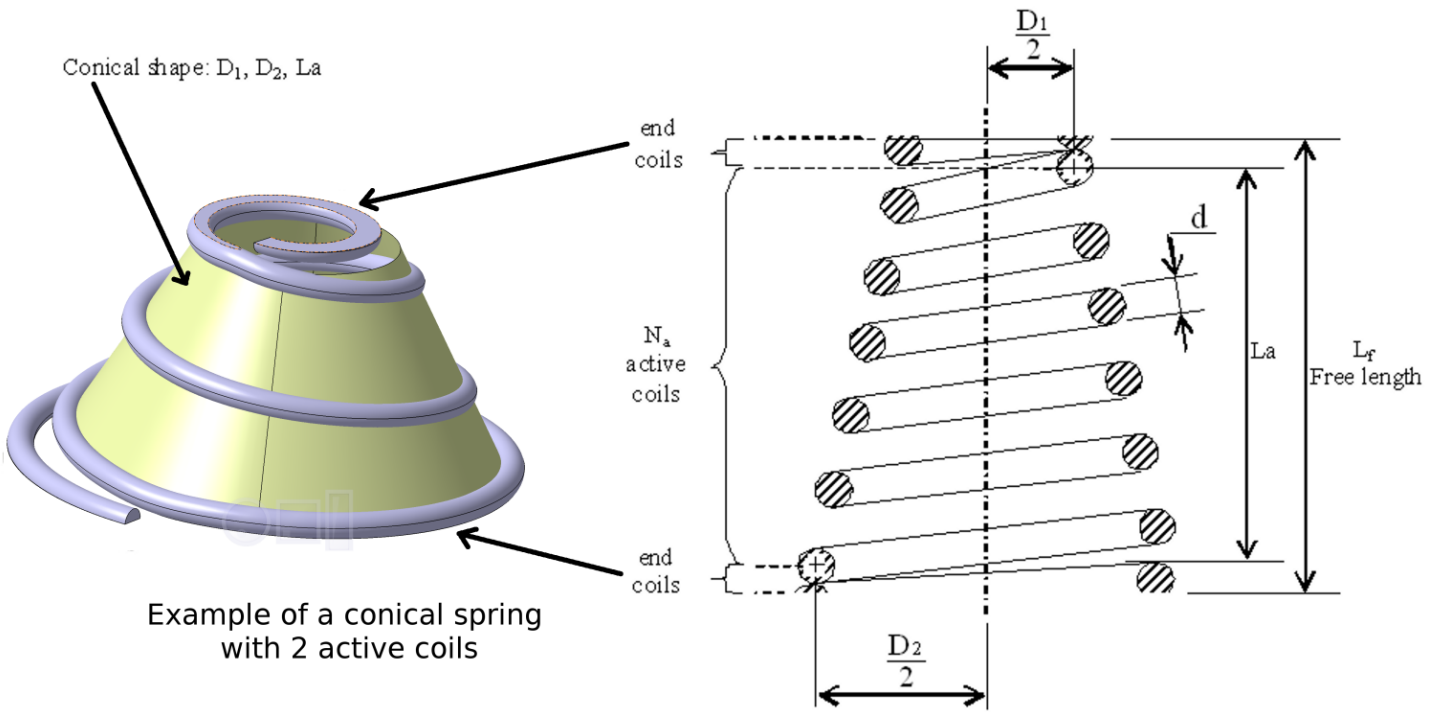

**Fig. 1 Parameters of a conical spring**

## 2.2 HELIX GEOMETRY

The geometry of the conical helix can be defined as follows:

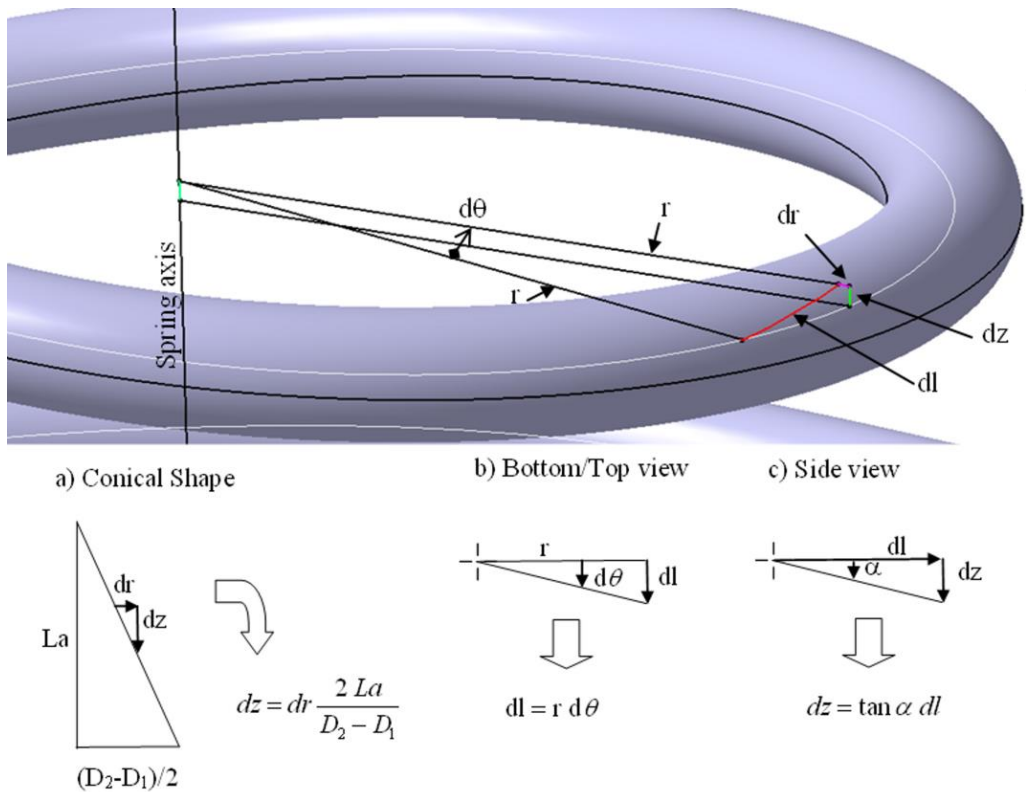

a) Conical Shape     b) Bottom/Top view     c) Side view

$$dz = dr \frac{2 La}{D_2 - D_1}$$

$$dl = r\, d\theta$$

$$dz = \tan \alpha \, dl$$

**Fig. 2 Parameterization of the helix**

Those relations enable us to define the global equation of the spiral:

$$\theta(R) = \int_{\frac{D_1}{2}}^{R} d\theta = \int_{\frac{D_1}{2}}^{R} \frac{dl}{r} = \int_{\frac{D_1}{2}}^{R} \frac{dz}{r \tan[\alpha(r)]} = \int_{\frac{D_1}{2}}^{R} \frac{2 La\, dr}{(D_2 - D_1) r \tan[\alpha(r)]} \quad (1)$$

In this equation, the helix angle $\alpha$ is a function of the current radius r and depends on the type of conical spring considered.

If the spring can telescope fully, the next equation must be satisfied for any value of $\theta$:

$$R(\theta + 2\pi) - R(\theta) \geq d \quad (2)$$

$$\text{With } 0 \leq \theta \leq 2\pi(N_a - 1)$$

## 2.3 ELEMENTARY CALCULATIONS

To determine the elementary calculations, we exploit the common assumptions for spring design. Firstly, the external load is considered as a pure axial load and, moreover, only the torque effect is considered in the calculations. The quality of this assumption depends directly on the geometry of the end coils. Secondly, it is assumed that, upon deflection, the generic point of the helix moves vertically (angle $\theta$ is maintained). This implies that the length of the wire shortens upon deformation, which is only acceptable for small helix angles. Finally, the effect of curvature is neglected; the elementary deflection is that of a straight bar in torsion.

The stress in an elementary coil defined by its radius r and subjected to a load P can be calculated using the formula proposed by Wahl [4] (uncorrected stress of a coil that is free to deflect):

$$\tau = \frac{16 r P}{\pi d^3} \quad (3)$$

Similarly, the elementary flexibility is:

$$F_e = \frac{64\, r^3}{G d^4} \quad (4)$$

and the elementary axial pitch is:

$$p_e = 2\pi r \tan\alpha \quad (5)$$

## 3. CHARACTERISTICS OF A CONICAL SPRING WITH A CONSTANT HELIX ANGLE

### 3.1 INITIAL GEOMETRY AND INITIAL STIFFNESS

Eq. (1) shows that a conical compression spring with a constant helix angle leads to a logarithmic spiral:

$$\theta(R) = \frac{2La}{(D_2 - D_1)\tan\alpha} \int_{\frac{D_1}{2}}^{R} \frac{dr}{r} = \frac{2La}{(D_2 - D_1)\tan\alpha} Ln\left(\frac{2R}{D_1}\right) \quad (6)$$

Knowing that $\theta\left(\frac{D_2}{2}\right) = 2\pi N_a$, Eq. (6) enables $\tan\alpha$ to be calculated:

$$\tan\alpha = \frac{La}{(D_2 - D_1)\pi N_a} Ln\left(\frac{D_2}{D_1}\right) \quad (7)$$

Now Eq. (4) can be reversed:

$$R(\theta) = \frac{D_1}{2} e^{\frac{\theta}{2\pi N_a} Ln\left(\frac{D_2}{D_1}\right)} \quad (8)$$

The initial flexibility can also be determined.

$$F = \frac{64}{Gd^4}\int_0^{N_a} R^3 dn = \frac{4 D_1^3}{\pi Gd^4}\int_0^{2\pi N_a}\left[e^{\frac{\theta}{2\pi N_a}Ln\left(\frac{D_2}{D_1}\right)}\right]^3 d\theta = \frac{8 N_a (D_2^3 - D_1^3)}{3Gd^4 Ln\left(\frac{D_2}{D_1}\right)} \quad (9)$$

It is important to note that, at this step, the fact that the conical spring is expected to fully telescope has not yet been exploited. This means that Eq. (9) represents the initial flexibility of any conical spring with a constant helix angle. It can also be seen that the flexibility of a conical spring with a constant angle is different from the flexibility obtained for conical springs with a constant pitch:

$$F = \frac{2 N_a (D_1^2 + D_2^2)(D_1 + D_2)}{Gd^4} \quad [23]$$

When using a conical spring, it is thus very important for researchers and designers to clearly explain not only the conical shape but also the way the active coils are distributed.

### 3.2 DEFLECTION OF A FULLY TELESCOPING CONICAL SPRING WITH A CONSTANT ANGLE

Such a spring must satisfy Eq. (2) near $D_1$. This leads it to satisfy:

$$\frac{D_1}{2} e^{\frac{1}{N_a}Ln\left(\frac{D_2}{D_1}\right)} - \frac{D_1}{2} \geq d \quad (10)$$

For a given load P, the limit between the coils that are free to deflect and the coils that are at solid is obtained for $p_e = F_e P$. Thus, combining Eq. (4) and (5), the associated diameter can be defined:

$$D_L = \sqrt{\frac{\pi \tan\alpha \, Gd^4}{8P}} \quad (11)$$

Now the transition load $P_T$ and the maximum load $P_M$ can be calculated. $P_T$ is the load from which the spring starts to come to solid (when $D_L = D_2$). $P_M$ is the load when all the coils come to solid (when $D_L = D_1$).

$$P_T = \frac{\pi \tan\alpha \, Gd^4}{8 D_2^2}$$

$$P_M = \frac{\pi \tan\alpha \, Gd^4}{8 D_1^2}$$

For a load lower than $P_T$, the deflection curve is linear, thus Eq. (23) gives:

$$\delta(P \leq P_T) = F \, P = \frac{8 N_a (D_2^3 - D_1^3)P}{3Gd^4 Ln\left(\frac{D_2}{D_1}\right)}$$

For a load between $P_T$ and $P_M$, the deflection of the spring can be expressed as the addition of two deflections: the deflection of the part of the spring that is free to deflect (from $D_1$ to $D_L$) and the deflection of the part that is at solid (from $D_L$ to $D_2$).
Thus

$$\delta(P_T \leq P \leq P_M) = \delta_1 + \delta_2 \text{ with}$$

$$\delta_1 = \frac{8 N_L (D_L^3 - D_1^3)P}{3Gd^4 Ln\left(\frac{D_L}{D_1}\right)} \quad \text{with } N_L = N_a \frac{Ln\left(\frac{D_L}{D_1}\right)}{Ln\left(\frac{D_2}{D_1}\right)} \quad \text{thus } \delta_1 = \frac{8 N_a (D_L^3 - D_1^3)P}{3Gd^4 Ln\left(\frac{D_2}{D_1}\right)}$$

$$\delta_2 = La\frac{(D_2 - D_L)}{(D_2 - D_1)}$$

## 4. CONICAL SPRING WITH CONSTANT STRESS AT SOLID
## 4.1 INITIAL GEOMETRY AND INITIAL FLEXIBILITY

An elementary coil situated at a radius r and subjected to a load P induces an elementary deflection (when it is free to deflect) of:

$$d\delta = P F_e\, dn = \frac{64\, P\, r^3}{G d^4} dn \quad (12)$$

When the elementary coil comes to solid, the following equation is obtained:

$$\tan\alpha = \frac{d\delta}{\pi 2 r dn} \quad (13)$$

Combining Eqs. (13), (12) and (3) enables us to calculate the value of the helix angle.

$$\tan\alpha = \frac{2\tau\, r}{G d} \quad (14)$$

Thus Eq. (1) becomes:

$$\theta(R) = \int_{\frac{D_1}{2}}^{R} \frac{2\, La\, dr}{(D_2 - D_1) r \tan\alpha} = \frac{G\, d\, La}{\tau_i (D_2 - D_1)} \left( \frac{2}{D_1} - \frac{1}{R} \right) \quad (15)$$

Where $\tau_i$ represents the constant stress at solid.

But we know that $\theta(\frac{D_2}{2}) = 2\pi N_a$ and thus the constant value of $\tau_i$ can be calculated:

$$\tau_i = \frac{G\, d\, La}{\pi\, N_a\, D_2\, D_1} \quad (16)$$

Eq. (15) and (16) can be combined:

$$\theta(R) = 2\pi N_a \frac{\frac{1}{D_1} - \frac{1}{2R}}{\frac{1}{D_1} - \frac{1}{D_2}} \quad (17)$$

And Eq. (17) can be reversed:

$$R(\theta) = \frac{1}{\frac{2}{D_1} - \frac{D_2 - D_1}{D_2 D_1 \pi N_a} \theta} \quad (18)$$

Eq. (18) is used to calculate the initial flexibility.

$$F = \frac{32}{\pi G d^4} \int_0^{2\pi N_a} R^3 d\theta = \frac{4 N_a D_1 D_2 (D_1 + D_2)}{G d^4} \quad (19)$$

## 4.2 DEFLECTION OF A FULLY TELESCOPING CONICAL SPRING WITH CONSTANT STRESS AT SOLID

Such a spring much satisfy Eqs. (2) and (18). This means it must satisfy:

$$\frac{D_1 D_2 N_a}{D_2 N_a - D_2 - D_1} - D_1 \geq 2d \quad (20)$$

The spring is expected to have a constant stress at solid $\tau_i$. This means that it is designed so that an elementary coil comes to solid when its stress is equal to $\tau_i$. Thus Eq. (3) can be used to calculate, for a given load P, the diameter that separates the part of the spring that is free to deflect from the part that is already at solid:

$$D_L = \frac{\pi \tau_i d^3}{8P} \quad (21)$$

Now the transition load $P_T$ and the maximum load $P_M$ can be calculated using Eqs. (16) and (21). $P_T$ is still the load from which the spring starts to come to solid (when $D_L=D_2$). $P_M$ is the load when all the coils come to solid (when $D_L=D_1$).

$$P_T = \frac{\pi \tau_i d^3}{8D_2} = \frac{La\, G\, d^4}{8 D_1 D_2^2 N_a}$$

$$P_M = \frac{\pi \tau_i G d^3}{8D_1} = \frac{La\, G\, d^4}{8 D_2 D_1^2 N_a}$$

For a load lower than $P_T$, the deflection curve is linear. Thus Eq. (19) gives:

$$\delta(P \leq P_T) = F\, P = \frac{4 N_a D_1 D_2 (D_1 + D_2) P}{G d^4}$$

For a load between $P_T$ and $P_M$, the deflection of the spring can once again be expressed as the addition of two deflections: the deflection of the part of the spring that is free to deflect (from $D_1$ to $D_L$) and the deflection of the part that is at solid (from $D_L$ to $D_2$). Thus

$$\delta(P_T \leq P \leq P_M) = \delta_1 + \delta_2 \quad \text{with}$$

$$\delta_1 = \frac{4 N_L D_1 D_L (D_1 + D_L) P}{G d^4} \quad \text{with} \quad N_L = N_a \frac{\frac{1}{D_1} - \frac{1}{D_L}}{\frac{1}{D_1} - \frac{1}{D_2}}$$

$$\delta_2 = La \frac{(D_2 - D_L)}{(D_2 - D_1)}$$

$D_L$ is calculated using Eq. (21).

## 5. CONICAL SPRING WITH LINEAR BEHAVIOR

### 5.1 INITIAL GEOMETRY AND FLEXIBILITY

To obtain fully linear behavior, any elementary coil situated at a radius r should come to solid for the same load $P_M$. The elementary deflection thus has to be equal to the elementary axial pitch. Thus Eqs. (4) and (5) give:

$$\tan \alpha = \frac{32 P_M r^2}{\pi G d^4} \quad (22)$$

Eq. (1) then becomes:

$$\theta(R) = \int_{\frac{D_1}{2}}^{R} \frac{2\, La\, dr}{(D_2 - D_1) r \tan \alpha} = \frac{G d^4 La \pi}{32 (D_2 - D_1) P_M} \left( \frac{4}{D_1^2} - \frac{1}{R^2} \right) \quad (23)$$

But we still know that $\theta(\frac{D_2}{2}) = 2\pi N_a$. Thus, the constant value of $P_M$ can be calculated:

$$P_M = \frac{G d^4 La}{16 N_a (D_2 - D_1)} \left( \frac{1}{D_1^2} - \frac{1}{D_2^2} \right) \quad (24)$$

There is no transition load for this kind of conical spring.
Eq. (23) can be simplified by exploiting Eq. (24):

$$\theta(R) = 2\pi N_a \frac{\frac{1}{D_1^2} - \frac{1}{4R^2}}{\frac{1}{D_1^2} - \frac{1}{D_2^2}} \quad (25)$$

and Eq. (25) can be reversed:

$$R(\theta) = \frac{1}{2\sqrt{D_1^{-2} - \frac{D_1^{-2} - D_2^{-2}}{2\pi N_a}\theta}} \qquad (26)$$

Eq. (26) is used to calculate the flexibility.

$$F = \frac{32}{\pi G d^4} \int_0^{2\pi N_a} R^3 d\theta = \frac{16 N_a D_1^2 D_2^2}{G d^4 (D_1 + D_2)} \qquad (27)$$

Being fully linear, this kind of spring is the only one to offer a conical shape at any compression state. The other kinds of springs have a conical shape only when uncompressed and their shape is non-conical when they are compressed, even for conical springs with constant pitch.

## 6. CASE STUDY

### 6.1 INITIAL GEOMETRY

The following case study illustrates the work presented above. Various conical springs, all having the same design parameters are considered (see details in Table 1).

Table 1: Details of the parameters used in the case study

| | | |
|---|---|---|
| d | 6 | mm |
| $D_1$ | 45 | mm |
| $D_2$ | 100.00 | mm |
| La | 50 | mm |
| $N_a$ | 2 | |

Eqs. (8), (18), (26) give the details of the relation between R and θ. They enable us to compare the spirals obtained to the one for a conical spring with constant pitch. They all have the same number of turns (coils) and the same values of $D_1$ and $D_2$.

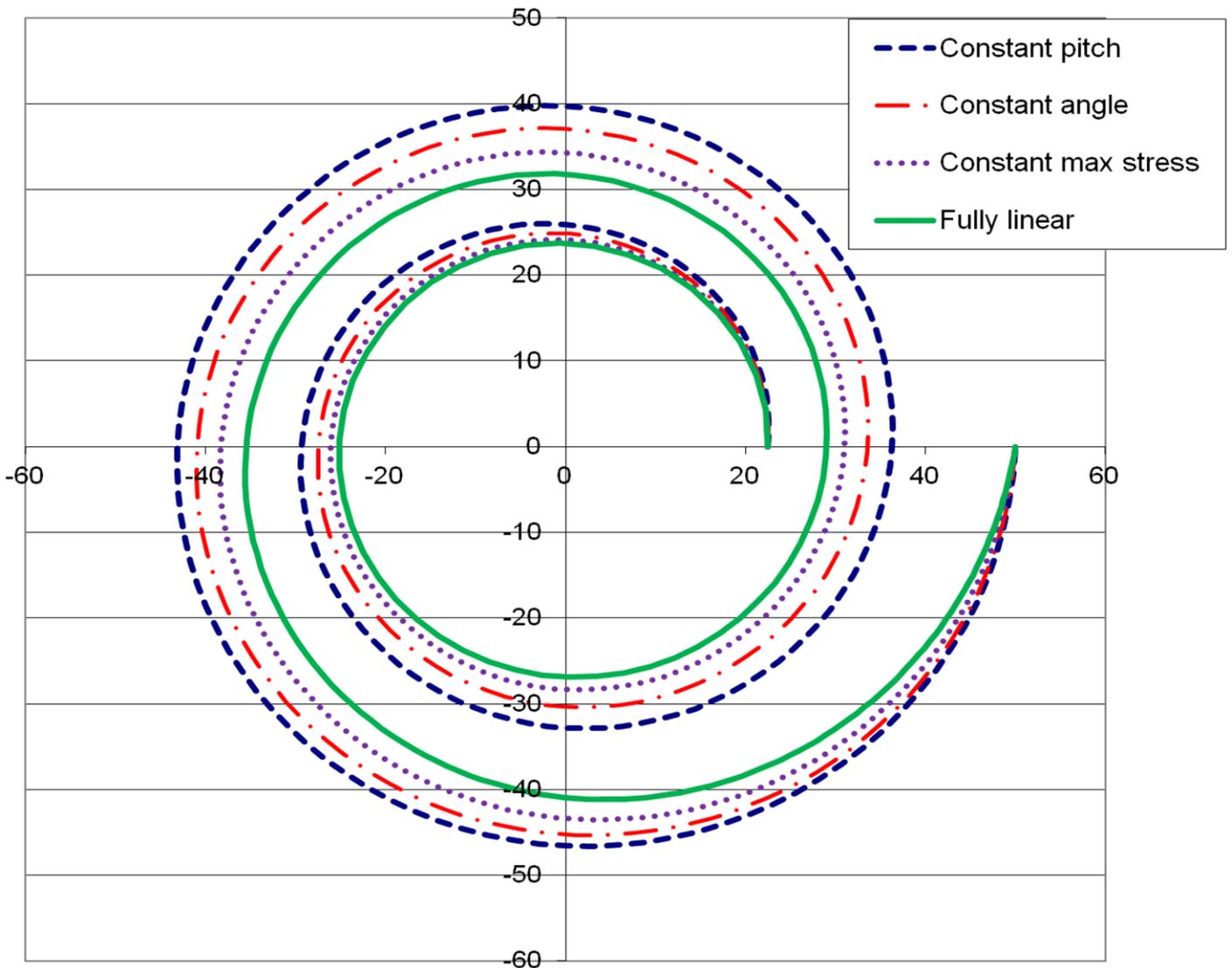

*Fig. 3 Detail of the helix of each conical spring*

It can be seen in Fig. 3 that, for the conical springs studied, the radial pitch is no longer constant but increases with radius. The fully linear conical spring is the one that has the smallest radial pitch. It will thus be the one for which the full telescoping property will induce the most design constraints.

**6.2 DEFLECTION OF FULLY TELESCOPING CONICAL SPRINGS**

All the conical springs comply with the data of table 1. They have the same free length and solid length. The equations detailing the deflection of conical springs are used to present the load/length curves for each one. The equations related to conical springs with constant pitch are presented by Rodriguez [23]. Fig. 4 was constructed by considering springs made of steel with G=81500 MPa.

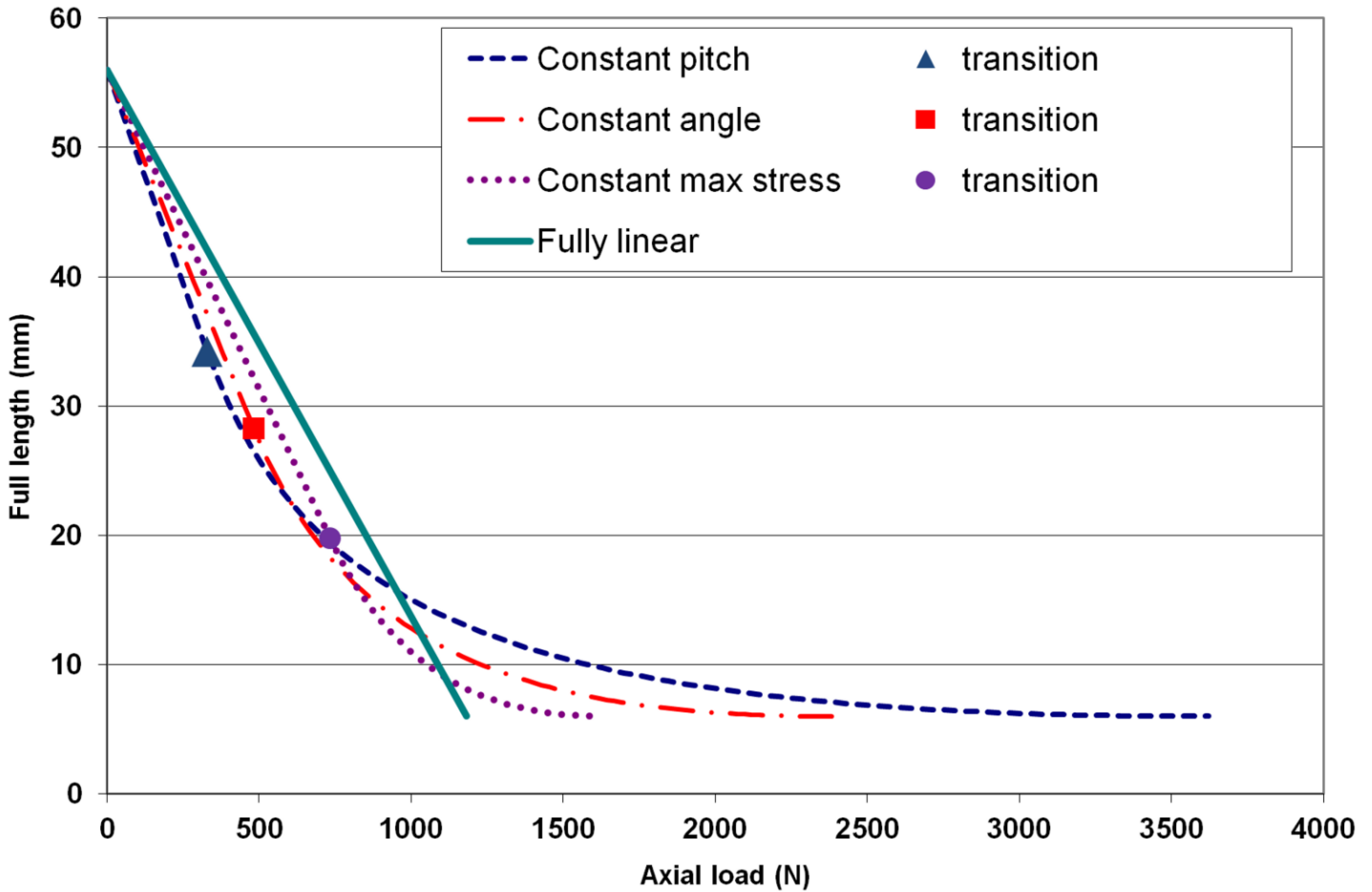

*Fig. 4 Load-length relations obtained*

We note that the proposed conical springs all have an initial flexibility that is lower than that of a conical spring with a constant pitch and, also, all have a lower maximum load (when compressed at solid). From a designer's point of view, it is very interesting to note the diversity of behaviors that are obtained by changing only the way the active coils are distributed along a same conical shape.

**6.3 EXPERIMENTAL STUDY**

In order to test the accuracy of the analytical approach, an experimental study was performed. Making conical springs out of steel with various helix shapes is not an easy matter. We could expect that numerically controlled coiling machines would be able to make such springs. However, adjusting the production process for each spring would require a long time and we need just one spring of each type for the study. For that reason, we decided to use a rapid prototyping technique.

Each spring was modeled using the CATIA Computer Aided Design software. Then the springs were manufactured in ABS plastic by means of Fused Deposition Modeling (on a Stratsys FDM 2000) as shown in Fig. 5.

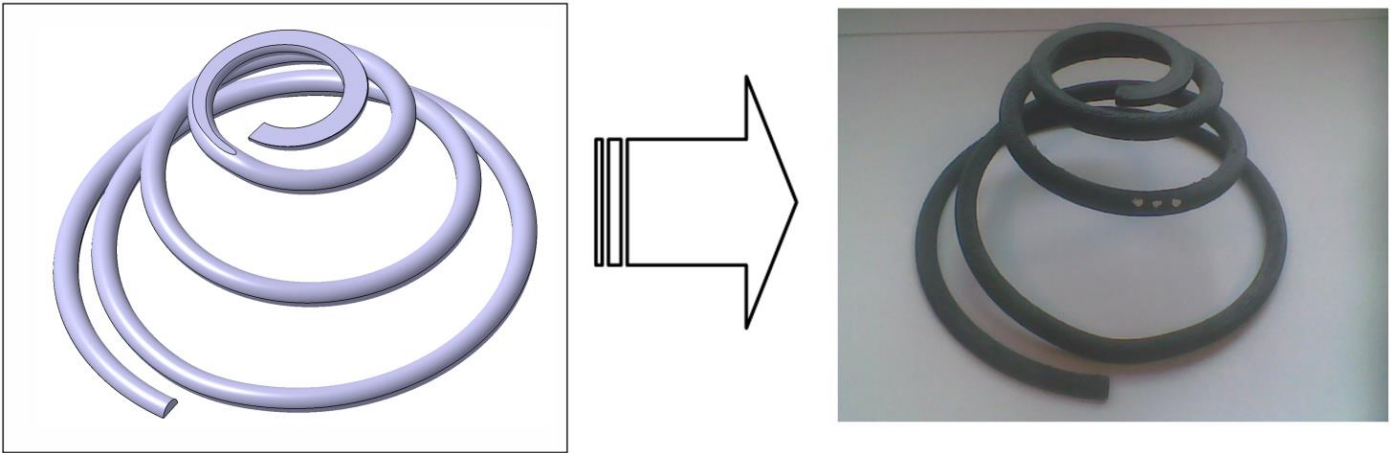

*Fig. 5 CAD model and geometry obtained*

In order to enable the springs to fully telescope to a length equal to the wire diameter, closed ground ends were used. They had an axial pitch of 0.5d and a radial pitch of 1.5d as shown in Fig. 6.

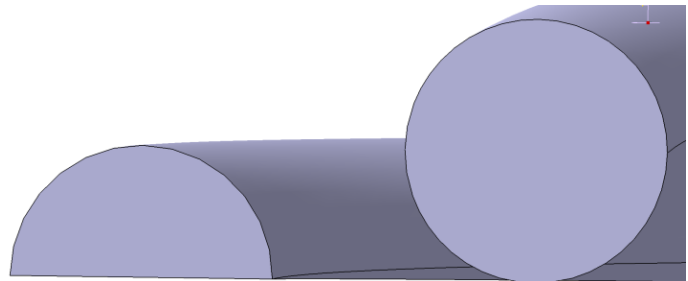

*Fig. 6 Detail of end coil design related to $D_2$*

In order to be able to compare experimental data and theoretical values, the torsional modulus of the wire formed by Fused Deposition Modeling needed to be evaluated. To do that, two cylindrical compression springs with the same wire diameter were built (see table 2). They had common closed ground ends.

Table 2: Details of the cylindrical springs

|       | Spring 1 | Spring 2 |     |
|-------|----------|----------|-----|
| d     | 6        | 6        | mm  |
| D     | 45       | 65.4     | mm  |
| L0    | 60       | 56       | mm  |
| $N_a$ | 2        | 2        |     |

The experimental tests on the two cylindrical springs determined the value of G:
G = 500 MPa. The comparison between theory and experiments is shown in Fig. 7.

As expected, the linear part of the experimental curves fits the theoretical ones perfectly.
Note that a gap appears for cylindrical springs when the length is near the solid length (see Fig. 7). For a given length, experimental data show a greater load than theory would predict. This is a well known issue for cylindrical springs and explains why such springs are never expected to be used near their solid length when linear behavior is required. This gap is mainly due to the fact that, as the axial load is not exactly in the spring axis, the contact of the active coils with the end coils does not occur at the same time everywhere. Now, the analytical formulae can be used and compared to the experimental results for the conical springs proposed. Four springs were tested. The results are presented in Fig. 8.

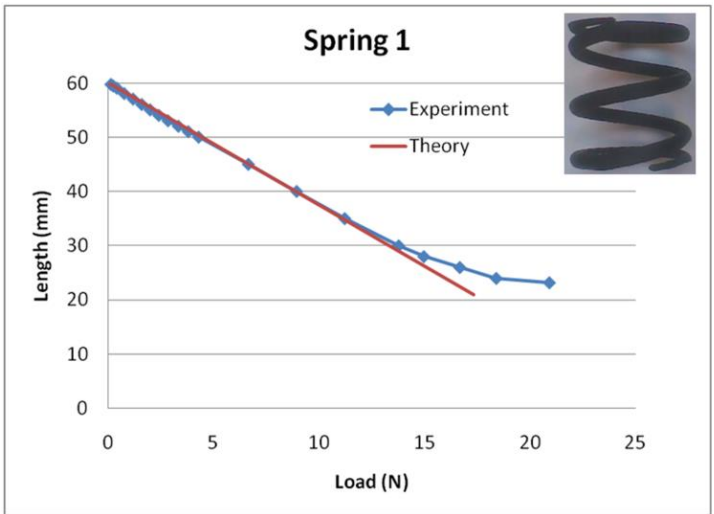
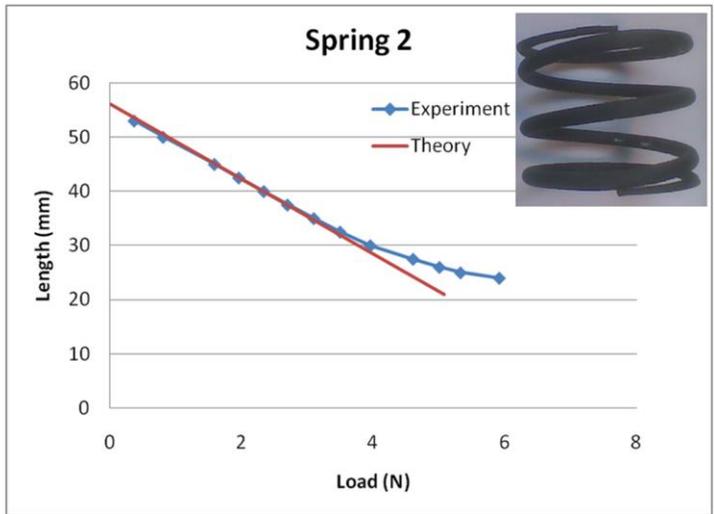

*Fig. 7 Experimental tests of cylindrical springs*

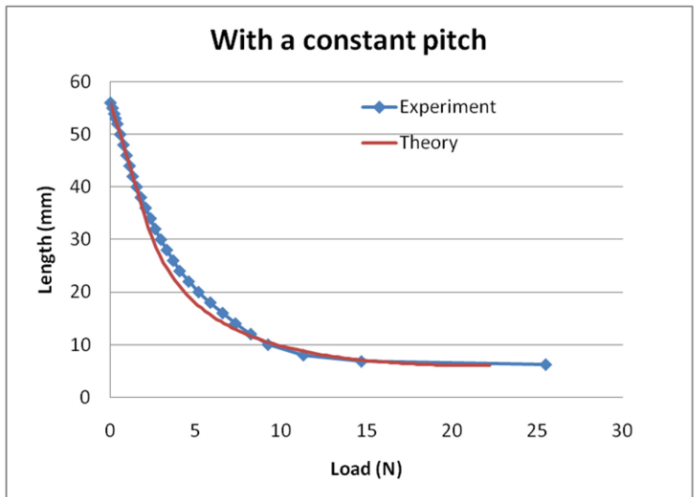
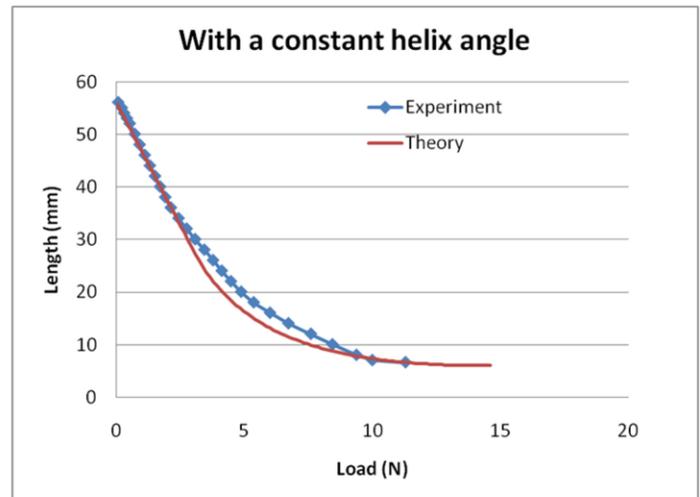
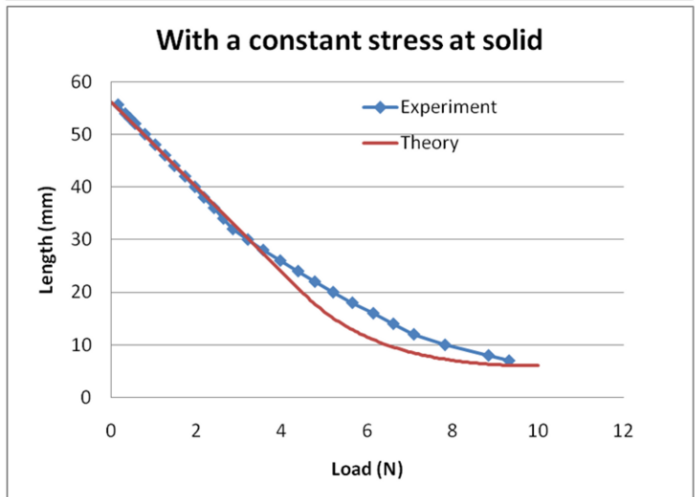
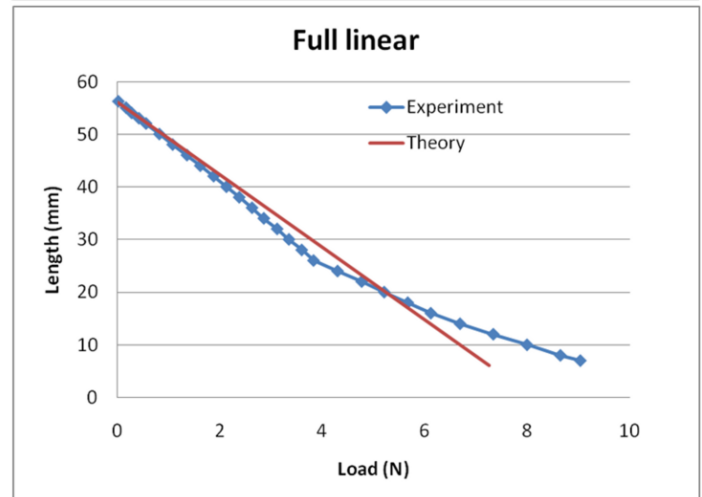

*Fig. 8 Experimental tests of conical springs*

For each kind of conical spring, the first part of the experimental Load/Length curve fits the theoretical curve perfectly. This proves that the bases of the theoretical approach are relevant and that the formulae for determining the initial rates are correct.

When each spring starts telescoping, a gap appears between the theoretical curve and the experimental one. This phenomenon is similar to the one that occurs for cylindrical springs. The experiments are driven by an axial displacement of the end coils. The load is transmitted to the spring mainly at two points, resulting in a non-axial load. This induces bending in the spring which is not yet considered. Nevertheless, it would be of great interest to be able to predict the Load/Length curve of telescoping springs more precisely near their solid length.

## 7. CONCLUSIONS

Most research papers that exploit conical springs focus only on conical springs with a constant pitch. In order to increase the range of possibilities, this paper has studied conical springs with other types of spirals projected on the conical shape.

The analytical study enabled us to define the theoretical geometry of the spiral in order to obtain a conical spring with a constant angle, with a constant maximum stress and with a fully linear load-length relation for fully telescoping springs.

Based on the spirals proposed, the corresponding initial flexibilities have been calculated using the common assumptions for springs. The formulae can be used for any kind of conical shape (whether the spring is able to telescope or not). The load-length relations have also been described in their non-linear range but only for fully telescoping springs.

Note that all the initial flexibility formulae (9), (19), (27) for conical springs lead to the standard equation for a cylindrical compression spring [4, 15] when $D_1$ tends to $D_2$.

Tests on conical springs made using Fused Deposition Modeling showed that all the analytical formulae proposed enable the initial rates to be determined with accuracy.

On the other hand, the theoretical formulae related to the nonlinear behaviors tend to under- estimate the load required to reach a given length. It would be of great interest to increase the accuracy of the predictions. Thus advanced finite element studies could be used to evaluate the effect of end coils, which is suspected to be the main cause of the gap. To reach the required accuracy, the studies would have to manage large displacements, contact between coils and contact between coils and the ground. Such finite element studies would be able to test several options for the end coil geometry and may help to find the most suitable ones.

Another source of improvement could be to precisely identify the loads (forces and moments) induced by the end coils as they are, and toperform another analytical study to determine the associated load-length relations.